\documentclass[aps,prb,twocolumn,superscriptaddress,10pt]{revtex4-2}
\usepackage{amsmath}
\usepackage{amssymb}
\usepackage{bm}
\usepackage{epsfig}
\usepackage{graphicx}
\usepackage{color}
\usepackage[unicode=true,colorlinks=true,citecolor=blue,urlcolor=blue]{hyperref} 
\usepackage[utf8]{inputenc}
\usepackage[english]{babel}
\usepackage{xcolor}
\usepackage[normalem]{ulem}

\newcommand{\e}{\mathrm{e}}

\newcommand{\ket}[1]{\left\vert #1 \right\rangle}
\newcommand{\dd}[2]{\frac{\mathrm{d}#1}{\mathrm{d}#2}}

\begin{document}
\newcommand{\thetitle}{Quantum beats of exciton-polarons in CsPbI\texorpdfstring{\textsubscript{3}}{3} perovskite nanocrystals}
\title{\thetitle}
\author{A.~V.~Trifonov}
\email[correspondence address: ]{artur.trifonov@tu-dortmund.de}
\affiliation{Experimentelle Physik 2, Technische Universit\"at Dortmund, 44227 Dortmund, Germany}
\affiliation{Spin Optics Laboratory, St. Petersburg State University, Peterhof, 198504 St. Petersburg, Russia}
\author{M.~O.~Nestoklon}
\author{M.-A.~Hollberg}
\author{S.~Grisard}
\author{D.~Kudlacik}
\affiliation{Experimentelle Physik 2, Technische Universit\"at Dortmund, 44227 Dortmund, Germany}
\author{E.~V.~Kolobkova}
\affiliation{ITMO University, 199034 St. Petersburg, Russia}
\affiliation{St. Petersburg State Institute of Technology, 190013 St. Petersburg , Russia}
\author{M.~S.~Kuznetsova}
\affiliation{Spin Optics Laboratory, St. Petersburg State University, Peterhof, 198504 St. Petersburg, Russia}
\author{S.~V.~Goupalov}
\affiliation{Department of Physics, Jackson State University, Jackson, 39217 Mississippi, USA}
\author{J.~M.~Kaspari}
\author{D.~E.~Reiter}
\affiliation{Condensed Matter Theory, Department of Physics, Technische Universit\"at Dortmund, 44227 Dortmund, Germany}
\author{D.~R.~Yakovlev}
\affiliation{Experimentelle Physik 2, Technische Universit\"at Dortmund, 44227 Dortmund, Germany}
\author{M.~Bayer}
\affiliation{Experimentelle Physik 2, Technische Universit\"at Dortmund, 44227 Dortmund, Germany}
\affiliation{Research Center FEMS, Technische Universit\"at Dortmund, 44227 Dortmund, Germany}
\author{I.~A.~Akimov}
\email[correspondence address: ]{ilja.akimov@tu-dortmund.de}
\affiliation{Experimentelle Physik 2, Technische Universit\"at Dortmund, 44227 Dortmund, Germany}

% \show\maketitle gives
%\@author@finish \title@column \titleblock@produce \suppressfloats [t]\let \and \relax \let \affiliation \@gobble \let \author \@gobble \let \@AAC@list \@empty \let \@AFF@list \@empty \let \@AFG@list \@empty \let \@AF@join \@AF@join@err or \let \email \@gobble \let \@address \@empty \let \maketitle \relax \let \thanks \@gobble \let \abstract \@undefined \let \endabstract \@undefined \titlepage@sw {\vfil \clearpage }{}.
%\makeatletter
%\@author@finish
%\let\thetitle\title@column
%\let\theaffillist\@AFF@list
%\let\theauthors\@AAC@list
%\makeatother

%  \textsf{\textbf{\Large Supporting Information \\  \thetitle}} \\
%  {\textsf{\theauthors}\par}
%  {\textit{\theaffillist}\par}

\begin{abstract}
Exciton–phonon interactions govern the energy level spectrum and thus the optical response in semiconductors. In this respect, lead-halide perovskite nanocrystals represent a unique system, for which the interaction with optical phonons is particularly strong, giving rise to a ladder of multiple exciton states which can be optically excited with femtosecond pulses. We establish a new regime of coherent exciton-polaron dynamics with exceptionally long coherence times ($T_2\approx 300$~ps) in an ensemble of CsPbI$_3$ nanocrystals embedded in a glass matrix. Using transient two-pulse photon echo at 2~K temperature, we observe quantum beats between the exciton-polaron states. Within a four-level model, we directly quantify the exciton–phonon coupling strength through the Huang–Rhys factors of $0.05\div 0.1$ and $0.02\div 0.04$ for low-energy optical phonons with energies of 3.2 and 5.1~meV, respectively. The pronounced size dependence of both coupling strengths and phonon lifetimes offers a path to tune the optical transitions between polaron states and to tailor the coherent optical dynamics in perovskite semiconductors for solid-state quantum technologies.
\end{abstract}

\date{\today}
\maketitle

\section{Introduction}

Lead halide perovskite semiconductors attract close attention due to their intriguing optical and electrical properties with huge potential for photovoltaic and light emitting  applications~\cite{Green2014,Sutherland2016}. These materials provide a unique playground for polaron physics in the solid state governed by the interaction of electronic excitations with the phonons of the crystal lattice~\cite{Yamada2022}. In particular, several studies indicate that low-energy optical phonons in halide perovskites have an important influence on the conductivity, while acoustic phonons have a minor effect~\cite{Even2013,Wright2016}. This is in contrast to conventional semiconductors such as GaAs, where the interaction of electrons with acoustic phonons is dominant at cryogenic temperatures~\cite{bookCardona,GantmacherLevinson}. Furthermore, the carrier mobility is determined not by a few specific optical phonons, but by a large multitude of modes due to the complex phonon spectrum~\cite{Ponce2019}. 

Perovskite nanocrystals (NC) belong to the class of quantum emitters with discrete energy level spectrum, large spectral tunability, and high quantum yield, which makes them attractive as single photon sources~\cite{Protesescu2015, Huang2017}. At cryogenic temperatures, the elementary optical excitation in NCs is an exciton (electron-hole pair) which possesses long optical and spin coherence times up to hundreds of ps, comparable to its radiative lifetime~\cite{Lv2021,Utzat2019,Tamarat2020}. Long-lived coherent excitons are often considered as qubits and are appealing for possible applications in quantum communication~\cite{Zhu2024}. However, the strong exciton-phonon interaction leads to formation of exciton-polarons which modifies the energy level spectrum. In particular, the photoluminescence spectra of single NCs show pronounced optical phonon replicas which confirm the substantial exciton-phonon coupling with Huang-Rhys factors in the range from 0.05 to 0.4 in lead halide perovskite NCs~\cite{Fu2017,Kanemitsu2021,Guilloux2023,Voisin2023,ZhuRaino2024}. Recently, electron-phonon interaction dominated by optical phonons with a similar Huang-Rhys factor of about 0.4 was demonstrated in CsPbBr$_3$ NCs~\cite{Iaru2021} and other inorganic lead halide perovskite NCs~\cite{Liao2019} using Raman spectroscopy. Compared to InGaAs quantum dots~~\cite{Bastard-1999, Janke-2005}, these values are significantly larger and comparable to those reported in II-VI nanostructures~\cite{Kelley-2019}.  A distinctive feature of perovskite nanocrystals is the existence of low-energy optical phonons with a strong coupling to phonons. This combination substantially modifies the exciton level spectrum in the vicinity of the ground state~\cite{Liu2023} and represents an attractive testbed for investigation of the quantum dynamics under resonant femtosecond excitation markedly different from previously studied systems.

Time-resolved studies can provide direct access to the exciton dynamics. It is well established that scattering on incoherent phonons with a thermal distribution causes loss of exciton coherence~\cite{Borri2001,Liu2021}. Yet, less is known about the quantum dynamics of exciton-polarons, i.e., the hybrid excitations of an exciton and a phonon. Recently, transient four-wave mixing and two-dimensional Fourier spectroscopy revealed the importance of coherent exciton-phonon coupling for the ground and excited (exciton and biexciton) states in lead halide perovskites ~\cite{Becker2018a,Seiler2019,Zhao2019,Yu2021}. However, most of these studies were limited to the case of short optical coherence of excitons ($\le100$~fs) and therefore reported on the coherent evolution of either the exciton or  the ground electronic state modulated by the frequencies of phonons with much slower decoherence. In other words, the lack of optical coherence restricted the observations to the phonon dynamics, while the genuine coherent evolution of the coupled exciton–phonon system has remained inaccessible. This scenario was also documented for bulk semiconductors, nanostructures and organic molecules~\cite{Mukamel1991,Schoenlein1993,Woggon2000,Mariette2002,Wigger2022}.  However, the opposite quantum dynamics limit, where pronounced quantum beats of exciton-polarons are expected, has not yet been uncovered. Furthermore, another general question arises,  connected to the limitations of coherent control for resonant excitation with femtosecond pulses, since time-resolved studies reported quite short coherence times up to 20~ps in lead halide perovskites~\cite{Becker2018a, Cundiff2021}, in contrast with photoluminescence data of single NCs~\cite{Lv2021,Utzat2019,Tamarat2020}. 

In this work we establish a new regime of coherent dynamics governed by exciton-polarons in NCs with an exceptionally long coherence time of the zero-phonon exciton ($T_{2,X} \approx 300$~ps) on the one hand and a discrete spectrum of spectrally narrow optical phonon modes on the other hand. The regime manifests in long-lived exciton-polaron quantum beats, which are detected by transient two-pulse photon echoes in an ensemble of CsPbI$_3$ nanocrystals embedded in a glass matrix at low temperature, $T=2$~K.  The oscillatory signal is dictated by optical phonons with low frequencies corresponding to $3.2$ and $5.1$~meV energy, and can be described by the solution of the Lindblad equations for a set of four-level systems that include a phonon ladder in the ground and excited exciton-polaron states. Different sizes, ranging from 10 to 12~nm, were selectively addressed within the same ensemble of nanocrystals by varying the laser photon energy. From the relative amplitude of oscillations we directly evaluate the strength of the exciton-phonon coupling with Huang–Rhys factors of 0.06 and 0.02 for optical phonons with 3.2~meV and 5.1~meV energy, respectively. Correspondingly, the electron–phonon coupling strength doubles reaching values of 0.12 and 0.04 as the NC size is reduced from 12 to 10~nm. The theoretical analysis of the size dependence shows that both the deformation potential and the Fr\"ohlich mechanism of electron-phonon interaction qualitatively agree with the experimental data in the regime of weak exciton confinement. Our study demonstrates that the dynamics of a pure quantum state following resonant optical excitation requires consideration of polaron effects, whose strength and coherence can be controlled by composition and size of the NCs. This opens up new possibilities for on-demand coherent control and coherent phonon generation in lead-halide perovskite NCs.

\section{Photon echo and long coherence of zero-phonon exciton}
\label{seq:ZP}

\begin{figure*}
\includegraphics[width=\linewidth]{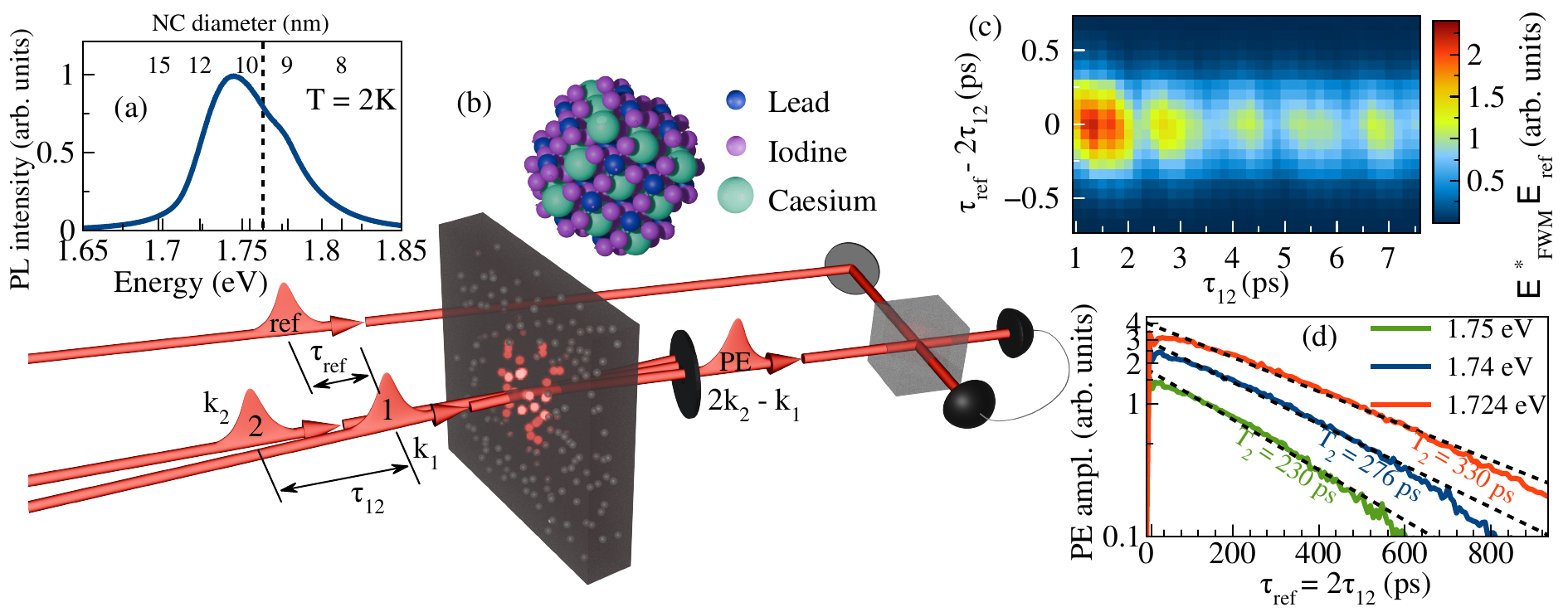}
\caption{{\bf Photon echoes in  CsPbI$_3$ nanocrystals.} 
(a) Typical photoluminescence (PL) spectrum for excitation with photon energy of 2.33 eV. Vertical dashed line indicates the highest photon energy used in the four-wave mixing FWM experiment. The scale on top corresponds to the NC diameter.
(b) Schematic representation of the experimental geometry, where the laser pulses hit the sample with wave vectors ${\bf k}_1$, ${\bf k}_2$. The photon echo is detected along the direction $2{\bf k}_2 - {\bf k}_1$ using a heterodyning technique by overlapping it with a reference pulse. Inset shows schematically a CsPbI$_3$ perovskite nanocrystal. (c) Two-dimensional plot of the FWM electric field amplitude $\mathcal{E}^*_{\rm FWM}$ as function of delay between the two pulses $\tau_{12}$ and the reference time $\tau_{\rm ref}$. The photon echo (PE) signal forms at the time $\tau_{\rm ref}=2\tau_{12}$. The amplitude of the PE shows oscillations during the initial evolution when $\tau_{12}$ is scanned. The signal is recorded in linearly co-polarized configuration.  Photon energy $h\nu=1.736$eV. (d) Decay of two-pulse photon echo amplitudes for excitation with different photon energies. Black dashed lines are fits with exponential functions, from which the labeld exciton coherence times $T_2$ are extracted. Temperature $T=2$~K.
  }
\label{Fig1}
\end{figure*}

The ensemble of CsPbI$_3$ NCs is synthesized in a fluorophosphate glass matrix by rapid cooling of a glass melt enriched with the materials needed for the perovskite crystallization. The details of synthesis are given in Ref.~\cite{Kolobkova2021}. The NC size is about $9\div 13$~nm in diameter (see Fig.~S6 in Ref.~\citenum{Nestoklon2023}). The low-temperature photoluminescence (PL) spectrum is dominated by a 60~meV broad band centered around 1.75~eV, as shown in Fig.~\ref{Fig1}(a). The broadening originates from fluctuations of the NC size, which we estimate to be in the order of 20\%. A single NC is schematically illustrated in the inset of Fig.~\ref{Fig1}(b). Note that the cubic crystal structure is shown for illustrative purposes as the actual crystallographic phase depends on the growth conditions and the size of the nanocrystals \cite{Marronnier2017,Marronnier2018,Wang2020,Yang2020}. 

We perform transient four-wave mixing (FWM) experiments in transmission geometry as shown schematically in Fig.~\ref{Fig1}(b). The NCs are resonantly excited with a sequence of two 100~fs laser-pulses with photon energies $h\nu <$1.76~eV in the low energy flank of the PL band. The sample temperature is kept at $T=2$~K. The electric field amplitude of the FWM signal $\mathcal{E}_{\rm FWM}(t)$ is resolved in time using heterodyne detection where the signal field is temporally overlapped with a strong reference pulse with  amplitude $\mathcal{E}_{\rm ref}(t-\tau_{\rm ref})$ (see Methods and Refs.~\citenum{Poltavtsev2016,Poltavtsev2020a,Poltavtsev2020,Poltavtsev2019} for details). Here, time $t=0$ corresponds to excitation with the first pulse, while $\tau_{\rm ref}$  is the delay of the reference pulse with respect to the first pulse in the excitation sequence. The resulting FWM signal is shown in Fig.~\ref{Fig1}(c), which gives a two-dimensional plot of $\mathcal{E}_{\rm FWM}$ as function of the delay time between the first and second pulse $\tau_{12}$ (vertical axis) and the reference delay time $\tau_{\rm ref}$ (horizontal axis). The FWM signal demonstrates the expected peak centered at time $\tau_{\rm ref}=2\tau_{12}$, corresponding to emission of a photon echo (PE) from the NC ensemble~\cite{Allen-Eberly-book}. This behavior arises from the inhomogeneous broadening of the optical transitions resulting from fluctuations of the NC size as demonstrated for similar halide perovskite NCs~\cite{Becker2018a, Cundiff2021}. Interestingly, during the first tens of ps the PE amplitude shows pronounced high frequency oscillations with increasing $\tau_{12}$ which we will discuss below. 

On a longer time scale, the amplitude of the photon echo decays exponentially with increasing delay time $\tau_{12}$ as shown in Fig.~\ref{Fig1}(d). Using exponential decay fits, shown with the dashed lines, we obtain $T_2=230 \div 330$~ps for photon energies in the range $1.724 \div 1.75$ eV, showing a weak decrease with increasing $h\nu$. The population relaxation time $T_1$, measured in three-pulse experiments, is found to be $T_1=800 \div 850$~ps, and is comparable to the exciton lifetime measured by time-resolved PL on the same sample. Therefore, we conclude that the signal at $\tau_{\rm ref} > 50$~ps is due to the zero-phonon exciton transition with long-lived optical coherence time $T_2$ and exciton recombination lifetime $T_1$. If the coherent dynamics were governed only by the population decay we would expect the relation $T_2 = 2 T_1$ to hold, which is the case for excitons in self-assembled InGaAs quantum dots at $T=2$~K~\cite{Kosarev2022}. Here, we observe a different situation, where elastic scattering processes (”pure dephasing”) with a decay constant of $T_{\rm p} = [1/T_2 - 1/(2T_1)]^{-1}=330$~ps mainly govern the exciton coherence.  Nevertheless, to the best of our knowledge, the homogeneous linewidth of the zero-phonon exciton of $\Gamma_2=2\hbar/T_2 = 4.8~\mu$eV demonstrated here has a record low value for perovskite nanocrystals~\cite{Lv2021,Utzat2019,Tamarat2020}.

\section{Coherent exciton-phonon dynamics}

\begin{figure*} 
\includegraphics[width=0.7\linewidth]{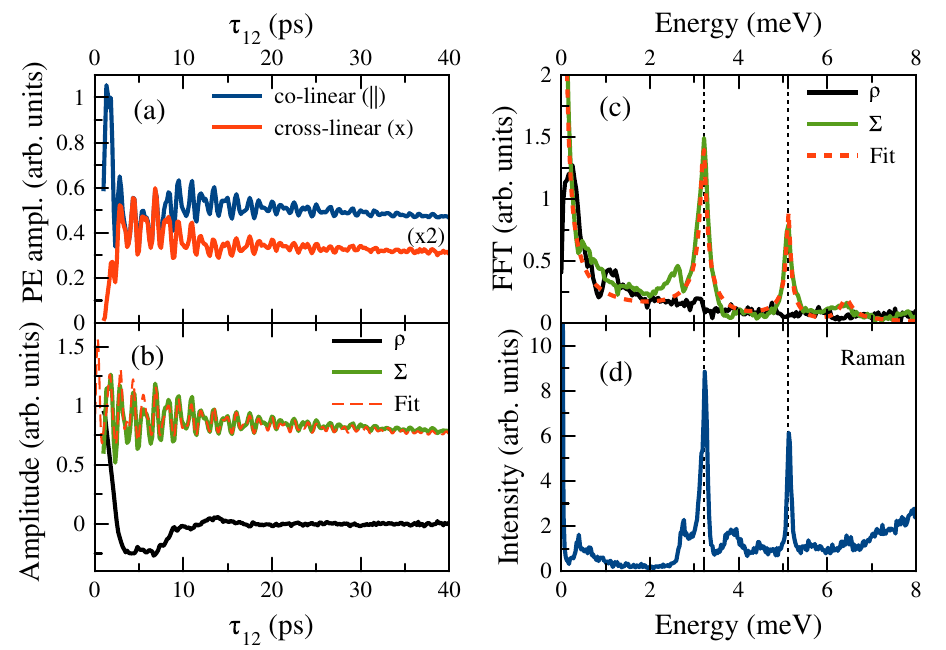}
\caption{{\bf Quantum beats of exciton-polarons.} (a)  Initial range of the PE dynamics in the co-linear ($\|$) and cross-linear ($\times$) polarization configurations shown with blue and red lines, respectively. Photon energy $h\nu=1.746$~eV. The amplitude of the $\times$ signal is multiplied by two for clarity. (b) Dynamics of $\rho$ (black) and $\Sigma$ (green) as defined by Eqs.~\ref{eq:rho} and \ref{eq:Sigma}, respectively. The dashed red curve is a fit using the polaron model with Eq.~\ref{eq:PhononEqs} with the following parameters for two phonon modes: $\hbar\Omega_1 = 3.2$~meV, $S_{\mathrm{HR1}}=0.097$, $\tau_{\mathrm{ph1}}=5.1$~ps; $\hbar\Omega_2 = 5.1$~meV, $S_{\mathrm{HR2}}=0.032$, $\tau_{\mathrm{ph2}}=10$~ps.
(c) Fast Fourier transform (FFT) amplitude spectra of $\rho$ (black), $\Sigma$ (green), and the fit curves in panel (b). Vertical dashed lines mark the peaks corresponding to optical phonons. (d) Raman spectrum measured at photon energy {$h\nu=1.734$~eV.} Vertical dashed lines indicate the positions of peaks corresponding to the optically active optical phonon modes that couple most strongly to the exciton.
}
\label{Fig:Osci}
\end{figure*}

As follows from Fig.~\ref{Fig1}(c) for short delay times $\tau_{12} \lesssim $ 10~ps, oscillations of the PE signal with multiple frequencies are observed. For deeper insight into the origin of the oscillations, we analyze polarization-resolved PE signals. These results are summarized in Fig.~\ref{Fig:Osci}. We employ two configurations, where the first and second pulses are linearly co- or cross- polarized, while the detection polarization in all cases coincides with that of the first pulse (see Methods). The PE amplitudes $A_{||}$ and $A_{\times}$ as function of $\tau_{12}$ in co-($\|$) and cross-($\times$) polarized configurations are shown in Fig.~\ref{Fig:Osci}(a). The oscillations can be grouped into two categories: slow oscillations with a period of 1--10~ps and fast oscillations with a period of less than 1~ps. 

It is evident that the fast oscillations in the co-polarized configuration are in phase and nearly identical for both curves. We will show below that these oscillations correspond to quantum beats between exciton-polaron states. By contrast, the slow oscillations are out of phase. The co-polarized $A_{\|}$-signal starts at its maximum value, while the $A_{\times}$-signal starts from zero. Subsequently, a minimum of the slow oscillations in $\|$-configuration around $\tau_{12}=5$~ps corresponds to a maximum in the $\times$-configuration. This out-of-phase behavior is related to quantum beats between the orthogonally polarized bright exciton states~\cite{Cundiff2021}. In perovskite nanocrystals, the bright exciton fine structure comprises three linearly polarized states~\cite{Fu2017,Nestoklon2018}, split by the energies $\delta_1$ and $\delta_2$ as shown in~Fig.~\ref{Fig:Scheme}(a). This energy level scheme is similar to the fine structure splitting in self-assembled quantum dots \cite{Bayer1999,Bayer2002} and in II-VI colloidal NCs~\cite{Furis2006,Htoon2009,Sinito2014,Goupalov2025}. In order to isolate the oscillations due to the fine structure splitting in the data, we model the expected polarization dependence of the PE signal, accounting also for the random orientation of the nanocrystals (for details see Sec.~\ref{sec:si:FSS} of the Supplementary Information). We obtain the following equations for the amplitude of the phonon echo signal in two polarization configurations:
\begin{subequations}\label{eq:APEaver}
\begin{align} 
A_{\|} & = \left[ \frac{18}{15} + \sum_{i} \frac{4}{15} \cos\left( \delta_i\tau_{12} \right) e^{-\frac{2\tau_{12}}{t_{i}^{*}}} \right] \Psi_0(\tau_{12})\,,
\\
A_{\times} & = \left[ \frac{6}{15} - \sum_{i} \frac{2}{15} \cos\left( \delta_i\tau_{12} \right) e^{-\frac{2\tau_{12}}{t_{i}^{*}}} \right] \Psi_0(\tau_{12})\,,
\end{align}
\end{subequations}
where the index $i=1,2,3$ corresponds to the beats between the three optical transitions ($\delta_3=\delta_1+\delta_2$), and $t_i^{*}$ is the dephasing time of the beats caused by the dispersion of the splitting energies $\hbar\delta_i$ in the ensemble. The common factor $\Psi_0(\tau_{12})$ is the same for both polarization configurations and will be discussed below, see also Supplementary Information \ref{sec:si:FSS}. Following Eqs.~\eqref{eq:APEaver} it is possible to distinguish between quantum beats related to the exciton fine structure and other polarization insensitive contributions to the PE signal by introducing  
the polarization contrast 
\begin{equation} 
\rho = \frac{A_{\|} - 3A_{\times}}{A_{\|} + 2A_{\times}},
\label{eq:rho}
\end{equation}
and the polarization sum
\begin{equation}
\Sigma = \frac{A_{\|} +  2 A_{\times}}{2}= \Psi_0(\tau_{12})\, .
\label{eq:Sigma}
\end{equation}
It follows that the temporal evolution of $\rho$ exhibits oscillations at frequencies corresponding to the energy splittings $\delta_i$ of the exciton fine structure and is independent of $\Psi_0(\tau_{12})$. In contrast, the polarization sum $\Sigma$ is given by the intrinsic coherent dynamics $\Psi_0(\tau_{12})$ only.

Figure~\ref{Fig:Osci}(b) shows the dynamics of these quantities, calculated from the data in Fig.~\ref{Fig:Osci}(a). In full accord with our expectations we obtain that the high-frequency oscillations are absent in the dynamics of $\rho$, while the low-frequency oscillations are still present. On the other hand, the low-frequency oscillations vanish in the $\Sigma$ transient in contrast to the high-frequency components. This becomes even more clear from the fast Fourier transform (FFT) spectra of $\rho$ and $\Sigma$ shown in Fig.~\ref{Fig:Osci}(c). From the fit of the $\rho$ transient we evaluate $\delta_1=0.25$~meV and $\delta_2=0.55$~meV, which are in agreement with the exciton fine structure splitting evaluated by two-dimensional Fourier spectroscopy on CsPbI$_3$ NCs of similar size with a diameter of about 9~nm~\cite{Cundiff2021}. A detailed discussion of the fine exciton structure combining pump-probe and PE studies will be published elsewhere. In what follows, we focus on the dynamics of $\Psi_0(\tau_{12})$, which is independent of the spin level structure of the exciton.

The FFT spectrum of $\Sigma$, shown by the green line in Figure~\ref{Fig:Osci}(c), exhibits several spectrally narrow features. These features are consistent with the peaks in the Raman spectrum in Figure~\ref{Fig:Osci}(d). The latter originates from light scattering on optically active phonons. The features at energies below 0.5~meV are attributed to confined acoustic phonons \cite{Hardcort2025}.
Below we focus on the two most prominent features marked by the vertical dashed lines with energies of $\hbar \Omega_1 = 3.2$~meV and $\hbar \Omega_2 = 5.1$meV, corresponding to the energies of optically active optical phonons in the vicinity of the $\Gamma$ point. Indeed optical phonon modes with energies close to 3 and 5~meV in CsPbI$_3$ were observed in Ref.~\cite{Liao2019} and calculated \cite{Ponce2019}. Optical phonons with similar energies were also detected by Raman spectroscopy in other lead halide perovskites such as FAPbI$_3$~\cite{Fu2018,Ferreira2020}, CsPbBr$_3$~\cite{ZhuRaino2024}, and CsPbCl$_3$~\cite{Calistru1997,Guilloux2023}. Thus, we conclude that the high frequency THz oscillations in the two-pulse coherent optical response are due to the coherent evolution of the coupled, hybridized exciton-phonon system. 

%==============================================================================
\section{Exciton-polaron quantum beats}

\begin{figure}
\includegraphics[width=0.9\linewidth]{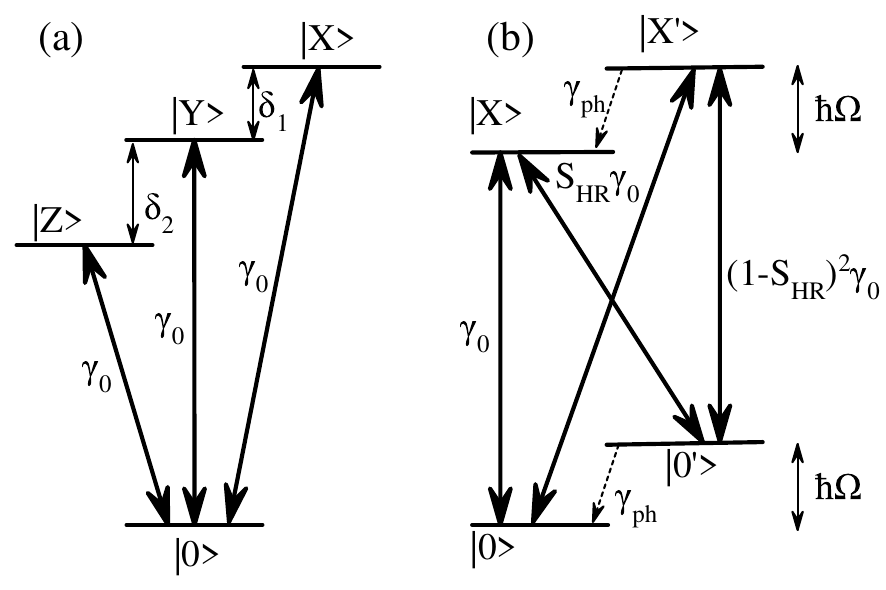}
\includegraphics[width=0.9\linewidth]{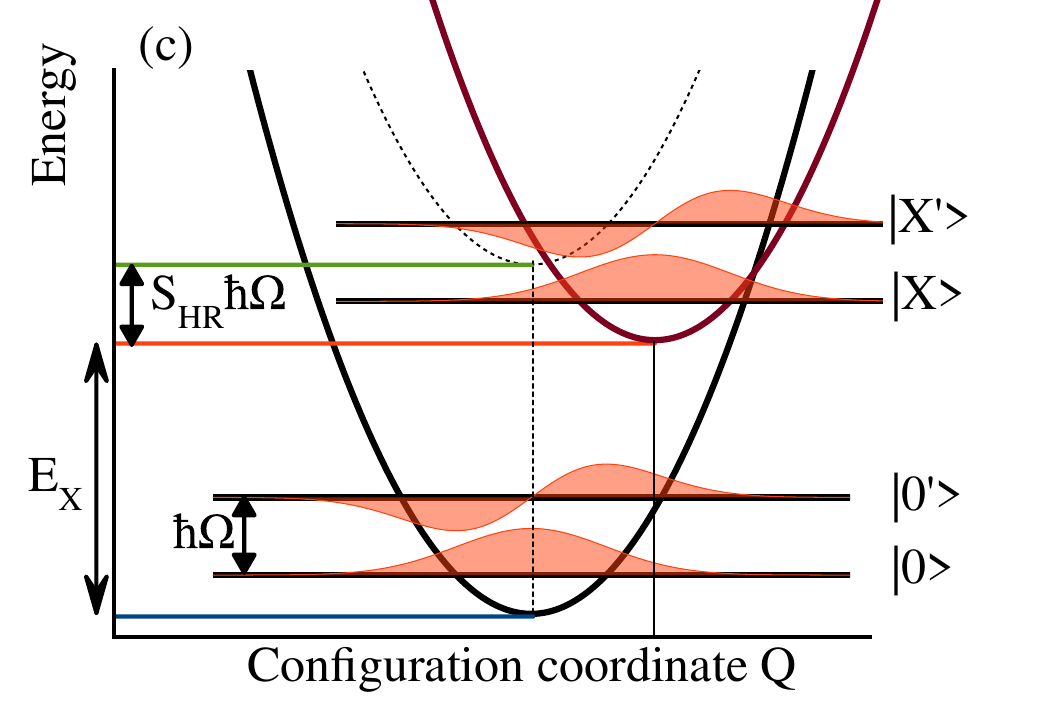}
\caption{{\bf Level schemes.}  (a) Energy levels of the bright exciton fine structure. (b) Energy level structure of the excitons interacting with optical phonons. We assume this level structure for each phonon mode, independent of the exciton polarization. The probability of the diagonal transitions is proportional to the Huang-Rhys factor $S_{\rm HR}$. $\gamma_{ph}$ is the phonon decay rate and $\gamma_0$ is the exciton decay rate, mostly determined by its oscillator strength. 
(c) Sketch of adiabatic potential of an exciton interacting with a particular phonon mode. The solid parabolas represent the potential for phonons for the ground state of the crystal (no exciton, bottom) and the excited state (with one exciton, top). The solid parabolas are separated by the energy difference $E_X$ corresponding to the energy of zero-phonon exciton transition. $\hbar \Omega$ denotes the energy of the optical phonon.}
\label{Fig:Scheme}
\end{figure}

In order to describe the quantum dynamics of the coupled exciton-phonon system we consider the structure of the optically excited energy levels in a single NC, which requires the involvement of polaron states. To that end, we introduce the four-level scheme shown in Fig.~\ref{Fig:Scheme}(b) which comprises the ground state $\ket{0}$, the phonon mode $\ket{0'}$, the exciton polaron state $\ket{X}$ (``exciton'') and the first excited exciton polaron state $\ket{X'}$ (see details in the Supplementary Information~\ref{sec:levels}). 
We neglect the exciton fine structure splitting to concentrate on the derivation of $\Psi_0(\tau_{12})$ in Eq.~\eqref{eq:APEaver}. Initially, we consider the exciton coupling to a single phonon mode in a NC. At the next stage, to compare with the experimental data, we perform summation over different phonon modes $\Omega_i$ where we neglect the interaction between them
\begin{equation}
  \Psi_0(\tau_{12}) = \sum_i \Psi_0(\tau_{12};\Omega_i)\,.
  \label{eq:PhononSum}
\end{equation}

The optical transitions between all energy levels are allowed using the same polarization. In the polaron model~\cite{bookAPY}, the transition probability amplitude is proportional to the product of the dipole matrix element and the overlap of the wavefunctions associated with vibronic modes shifted due to polaron formation (see Fig.~\ref{Fig:Scheme}(c) and Eq.~\ref{eq:gammaHR}), which depends on the strength of the exciton-phonon interaction given by the Huang-Rhys factor $S_{\rm HR}$ (Condon principle). The pulse duration is assumed to be short as compared with all other characteristic times in the system, i.e. we can assume the $\delta$-pulse limit. This is justified because the laser pulse duration is $\tau_d \approx 100$~fs (see Methods) and the fit to experimental data gives relaxation times in the order of 10~ps. The spectral width of the laser pulse $\sim h/\tau_d$ is about $15$~meV, which is larger than the phonon energies $\hbar\Omega_i$. Therefore, we assume that all four transitions between states $\ket{0}$, $\ket{0'}$ and $\ket{X}$, $\ket{X'}$ are covered by the spectral width of the laser. Note that a not too large Huang-Rhys factor means that states involving multiple phonons (i.e. $2\hbar\Omega_1$, etc.) may be neglected. However, the spectral width of the laser pulse is large enough to cover them as well. We neglect the possible difference of phonon energy in the ground and polaron state as we do not find corresponding frequency changes in our experimental data and in previous reports in other perovskites \cite{Fu2018,Ferreira2020}. Similar assumptions were made in the model developed for describing 2D spectroscopy in Ref.~\citenum{Seibt2013}.

In the polaron model, we trace the evolution of the density matrix components which contribute to the PE. The result arises from the coherent off-diagonal terms and all levels contribute to the result in first order in the Huang-Rhys factor $S_{\rm{HR}}$, which quantifies the exciton-phonon interaction. Solution of the Lindblad equation in Supplementary Information~\ref{sec:PEcalcul} gives the following expression for the amplitude of the phonon echo signal from a NC with single phonon mode $\Omega_i$
\begin{multline}\label{eq:PhononEqs}
  \Psi_0(\tau_{12}; \Omega_i) =  \e^{-(1+S_{\rm{HR}})\gamma_0 \tau_{12}} \Bigg[ 1 
 + S_{\rm{HR}} \e^{-(\gamma_{ph}-S_{\rm{HR}}\gamma_0) \tau_{12} }
\\
+ S_{\rm{HR}} \cos{(\Omega_i \tau_{12})} \e^{-\frac{\gamma_{ph} -2 S_{\rm{HR}}\gamma_0}{2} \tau_{12}}
\Big[  2 + \e^{- S_{\rm{HR}}\gamma_0 \tau_{12}} + \e^{-\gamma_{ph}\tau_{12}} \Big] \\
+ S_{\rm{HR}} \cos{(2 \Omega_i \tau_{12})} \e^{-(\gamma_{ph}-S_{\rm{HR}}\gamma_0) \tau_{12}} \Bigg] .
\end{multline}
Here $T_{2,0} = 1/\gamma_0$ is the coherence time associated with the zero-phonon optical transition, and $\tau_{ph} = 1/\gamma_{ph}$ is the lifetime of the optical phonon. In the simplified model we do not include other sources of decoherence. The analytical expression fully supports the experimental observations for $\Sigma=\Psi_0(\tau_{12})$ shown in Fig.~\ref{Fig:Osci}(b). The first term in Eq.~\ref{eq:PhononEqs} corresponds to the long-lived zero-phonon coherence which decays exponentially with the time $T_2=T_{2,0}/(1+S_{\rm HR})\sim$~300~ps, as evaluated in section ~\ref{seq:ZP}. The last two terms correspond to high frequency oscillations at the single and double frequency of the phonon mode $\Omega_i$, respectively. They appear due to quantum interference of excitations of different exciton-polaron states, i.e. due to quantum beats of exciton-polarons. These oscillations are superimposed on the long-lived signal and decay with a shorter time $\tau_{ph}$. The relative amplitude of the oscillatory signal and the long-lived plateau allows one to measure the Huang-Rhys factor $S_{\rm HR}$. 

Using Eqs.~\ref{eq:PhononSum} and \ref{eq:PhononEqs} for two independent phonon modes with $\hbar\Omega_1=3.2$~meV and $\hbar\Omega_2=5.1$~meV we obtain excellent agreement with the experimental data. The phonon frequencies are taken from the peak positions in the Fourier spectrum of Fig.~\ref{Fig:Osci}(c). The Huang-Rhys factors and phonon lifetimes are evaluated from the best fit to the experimentally measured transients, see Fig.~\ref{Fig:Osci}(b). We emphasize that in contrast to previous reports ~\cite{Mukamel1991,Schoenlein1993,Woggon2000,Mariette2002,Wigger2022}, our system demonstrates long-lived coherent dynamics, which is attributed to an exceptionally long zero-phonon exciton coherence ($\sim 300$~ps) and a relatively long phonon lifetime ($\sim 10$~ps). Here, we stress that the low-temperature regime with $T=2$~K is essential, as the coherence of the zero-phonon excitons rapidly vanishes with increasing temperature. Furthermore, the optical phonons each are represented by well-defined frequencies due to their flat dispersion around the $\Gamma$-point, as confirmed by the Raman spectrum in Fig.~\ref{Fig:Osci}(d), making CsPbI$_3$ NCs embedded in a glass matrix particularly interesting for investigating coupled exciton-phonon dynamics. The peak widths in the Raman spectrum can be recalculated into phonon lifetimes using the following relation  $\tau_{ph} = 2\hbar/\Delta E_R$~\cite{AkuLeh2005}, where $\Delta E_R$ is the full width at half maximum in energy units. This yields phonon lifetimes of approximately 7~ps and 11~ps for the 3.2 and 5.1~meV modes, respectively. These values are in excellent agreement with the decay times of the PE oscillatory signals. It should be noted that the Raman peaks in Fig.~\ref{Fig:Osci}(d) appear narrower than those in the FFT spectra in Fig.~\ref{Fig:Osci}(c), which originates from the difference between the power spectral and the amplitude spectral representations, respectively. 

The proposed polaron model (i) provides an intuitive picture of the microscopic processes involved, (ii) has a simple analytical solution, and (iii) accounts explicitly for the coherence decay through simple equations. Equation~(\ref{eq:PhononEqs}) gives a result that closely matches that in Refs.~\cite{Mukamel1991,Mukamel2004} and is in effect equivalent within the accuracy limits of the models. Note that a direct comparison between the results obtained by the two approaches is not straightforward: Eq.~\eqref{eq:PhononEqs} is obtained from the exact solution of the Lindblad equations for a four-level system, while Ref.~\citenum{Mukamel2004} considers the quasi-classical evolution of a complex system averaged over the phonon subsystem. For a detailed discussion and comparison of the two approaches see e.g. Ref.~\citenum{Boyanovsky2017}. 

%===============================================================================
\section{NC size dependence}
\label{seq:NCsize}

\begin{figure}
\includegraphics[width=\linewidth]{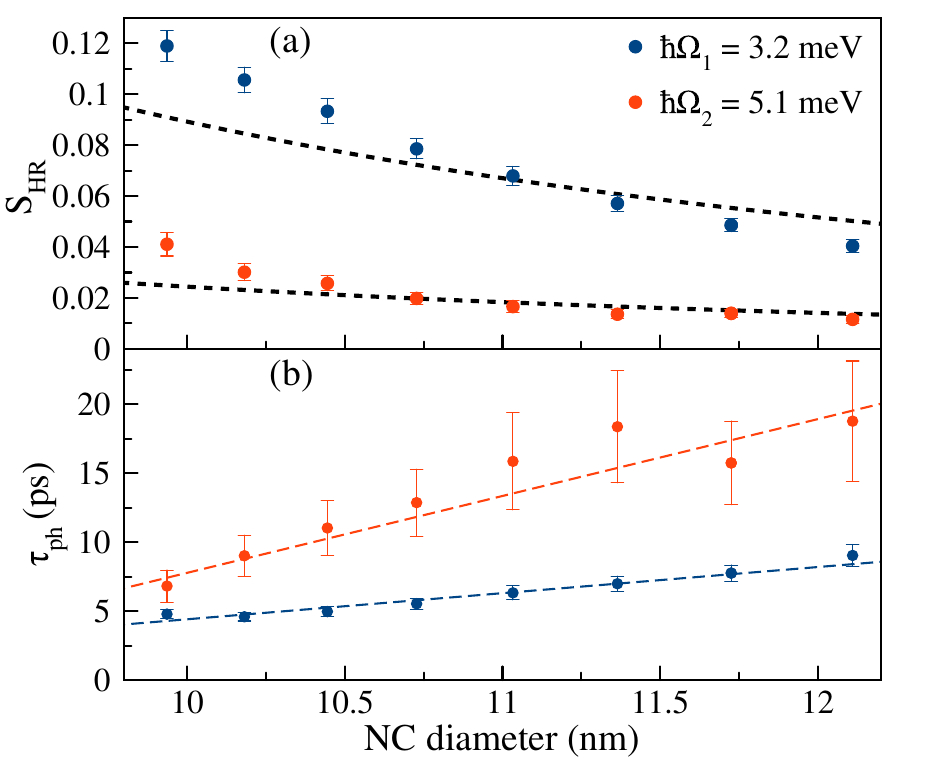}
\caption{{\bf NC size dependence.} 
Dependence of (a) the Huang-Rhys factor, $S_{\rm{HR}}$, and (b) the phonon lifetime, $\tau_{ph}$, for the two phonon modes with energies $\hbar\Omega_1=3.2$~meV (blue dots) and $\hbar\Omega_2=5.1$~meV (red dots) on NC diameter $a$. Dashed lines in (a) are fits proportional to $a^{-3}$. Dashed lines in (b) are guides to the eye.}
\label{Fig:HR}
\end{figure}

Above, we discussed the PE dependence measured for the fixed laser pulse photon energy of $h\nu = 1.734$~eV and explained how the phonon parameters can be evaluated from the PE signal. However, the sample under study contains NCs with different sizes which can be selectively excited by tuning the laser photon energy. We measured PE transients at photon energies in the range of $1.72\div 1.76$~eV. The photon energy can be recalculated into a NC diameter using the parameters from Ref.~\citenum{Nestoklon2023}, where the same sample was studied (sample \#3). Figures~\ref{Fig:HR}(a) and ~\ref{Fig:HR}(b) show the dependence of the evaluated Huang-Rhys factors $S_{\rm{HR}}$ and phonon lifetimes $\tau_{ph}$ as function of the NC diameter $a$, for phonon modes with energies $\hbar \Omega_1 = 3.2$~meV and $\hbar \Omega_2 = 5.1$~meV. We note that the frequencies of these modes do not depend on $a$ within the accuracy of the experiment. The Huang–Rhys factors for both phonon modes exhibit a clear size dependence, increasing with decreasing NC size. Correspondingly, the phonon lifetimes decrease with decreasing $a$. An increase in the electron–phonon coupling strength with decreasing NC size was previously reported for CdSe nanocrystals~\cite{Mittleman1994}, PbS quantum dots~\cite{Maisa2011}, and perovskite nanocrystals, both for optical and acoustic phonons~\cite{Kanemitsu2021, ZhuRaino2024}. However, to the best of our knowledge, direct measurements of the associated phonon relaxation times have not been reported. It is also important to note that, in contrast to previous studies, our PE experiments selectively probe the coherent dynamics of each individual phonon mode.

In Ref.~\citenum{ZhuRaino2024}, the dominant mechanism of electron-phonon coupling is assigned to the  optical deformation potential~\cite{Yazdani2024}. In this case, the size dependence of the interaction can be estimated from the phonon normalization condition \cite{GantmacherLevinson} which leads to $S_{\rm HR}\sim a^{-3}$. For completeness, let us discuss the size dependence of the interaction between charge carriers and optical phonons for the Fr\"ohlich mechanism following Takagahara~\cite{Takagahara1996}. In the original work, the interaction of charge carriers with phonons is rewritten as potential for an electron in the field induced by the phonon mode polarization ${\bf P}({\bf r})$. This potential is found from the Poisson equation $\bm{\nabla}^2 \varphi({\bf r})=4 \, \pi \bm{\nabla} \cdot {\bf P}({\bf r})$ which leads to
\begin{equation}\label{eq:phie}
\varphi({\bf r}_e)=\int d{\bf r} \, \frac{\bm{\nabla} \cdot {\bf P}({\bf r})}{|{\bf r}-{\bf r}_e|}=
-\int d{\bf r} \, {\bf P}({\bf r}) \cdot \bm{\nabla} \frac{1}{|{\bf r}-{\bf r}_e|} \,.
\end{equation} 
From Eq.~\eqref{eq:phie} the strength of electron-phonon interaction $\Delta_{e}\sim a^{-1}$ which leads to $S_{\rm HR}^{e}=\Delta_{e}^2/2\sim a^{-2}$. This estimate is valid also for excitons in the strong confinement regime.

For weakly confined excitons, the change of exciton energy $\Delta$ is proportional to the difference of the electrostatic potential for electron and hole $\varphi({\bf r}_e)-\varphi({\bf r}_h)$. It is found to be the sum of Eq.~\eqref{eq:phie} for electron and hole:
\begin{equation}
\label{Poisson}
\Delta \sim \varphi({\bf r}_e)-\varphi({\bf r}_h)=\int d{\bf r} \, {\bf P}({\bf r}) \cdot \bm{\nabla} \left( \frac{1}{|{\bf r}-{\bf r}_h|} - \frac{1}{|{\bf r}-{\bf r}_e|} \right) \,.
\end{equation}
In the weak confinement regime when the exciton Bohr radius $a_B$ is small compared with the NC diameter $a_B \ll a$,
\begin{equation}\label{expansion}
\frac{1}{|{\bf r}-{\bf r}_h|} % = \frac{1}{|{\bf r}-{\bf r}_e+{\bf r}_e-{\bf r}_h|} 
\approx \frac{1}{|{\bf r}-{\bf r}_e|} + ({\bf r}_e-{\bf r}_h) \cdot \bm{\nabla} \frac{1}{|{\bf r}-{\bf r}_e|} \,.
\end{equation}
Substituting Eq.~(\ref{expansion}) into Eq.~(\ref{Poisson}) we obtain
\[
\varphi({\bf r}_e)- \varphi({\bf r}_h) \approx \int d{\bf r} \, {\bf P}({\bf r}) \cdot \bm{\nabla} \left[ ({\bf r}_e-{\bf r}_h) \cdot \bm{\nabla} \frac{1}{|{\bf r}-{\bf r}_e|} \right] \,.
\]
As a result, in the weak confinement regime, the size dependence of this expression is estimated as
\[
\Delta \sim \varphi({\bf r}_e)- \varphi({\bf r}_h) \propto a^3 \cdot a^{-3/2} \cdot \frac{a_B}{a^3} \propto  a^{-3/2}.
\]
Here, the first size factor comes from the integral, the second one from the normalization of the phonon mode, and the last one from the gradient of the expression in the square brackets. As a result the Huang-Rhys factor scales as $S_{\rm HR} = \Delta^2/2 \sim a^{-3}$.

Thus, in the weak confinement regime both mechanisms, the deformation potential and the Fr\"ohlich interaction, lead to similar dependences of the Huang-Rhys factor on the NC size $S_{\rm HR} \sim a^{-3}$. In the strong confinement regime, the Fr\"ohlich interaction is expected to have a weaker size dependence, while the deformation potential mechanism should show the same scaling. The dashed lines in Fig.~\ref{Fig:HR} represent fits of the experimental data by the empirical relation $S_{\rm HR} \propto a^{-3}$. The experimental data in Fig.~\ref{Fig:HR}(a) even shows a steeper dependence of the Huang–Rhys factor on NC size, with pronounced deviation from the $a^{-3}$ dependence, particularly for small NCs. This observation suggests that further analysis is required for smaller NCs to identify the mechanisms of electron–phonon interaction in perovskite NCs.

%===============================================================================
\section{Discussion}

To summarize, we have revealed a new regime of coherent exciton-phonon dynamics following resonant optical excitation with  femtosecond pulses in an ensemble of CsPbI$_3$ nanocrystals embedded in a glass matrix. Quantum beats between exciton-polaron states manifest themselves in two-pulse photon echo experiments as oscillations with high frequencies corresponding to two optical phonon modes at 3.2 and 5.1~meV. A polarization resolved analysis of the photon echo signal has allowed us to unambiguously isolate the quantum beat signatures of exciton-polarons from slower beats arising from exciton fine-structure splittings. This unique regime of exciton-polaron quantum beats is established at the low temperature of 2~K, where an exceptionally long zero-phonon exciton coherence time is achieved ($T_2 \approx 300$~ps). The presence of spectrally narrow optical phonon modes with lifetimes of about 10~ps, combined with a relatively strong exciton-phonon coupling (Huang-Rhys factors up to 0.1), leads to a pronounced modulation of the photon echo signal at the phonon frequencies. Furthermore, the strong size dependence of the Huang–Rhys factor makes it possible to tune the strength of the optical transitions between different polaron states, offering a route to control the quantum dynamics for targeted applications.

We emphasize that the observed optical response markedly differs from previous studies in, e.g., III–V self-assembled quantum dots, where the dynamics are mainly governed by exciton–acoustic phonon coupling~\cite{Borri2001, Ramsay-2010, Grisard-2022}. There, the continuous phonon spectrum leads to a partial loss of coherence within the first few picoseconds after pulsed excitation. In our case no such behavior is observed in the temporal evolution of $\Psi_0(\tau_{12})$ [see Fig.~\ref{Fig:Osci}(b)]. Instead, the coherent dynamics is dominated by exciton–polaron states, i.e., interactions with discrete optical phonon modes, which can be consistently described within a four-level model for each phonon frequency. The desired quantum evolution can be achieved by tuning both the exciton–phonon coupling strength and, most importantly, the phonon lifetime, which limits the evolution of the pure quantum state following resonant excitation with a short optical pulse. Notably, the phonon lifetime exhibits a pronounced dependence on nanocrystal size. For the optical phonon mode with an energy of 5.1~meV, we observe an almost twofold increase of $\tau_{ph}$ as the size increases from 10 to 12~nm. 
In layered perovskites, phonon lifetimes on the order of $\sim 10$~ps were reported \cite{Zhang2023}, with values extended up to 75~ps \cite{Yang2024}, approaching the record-high $\sim200$~ps observed in bulk ZnO \cite{AkuLeh2005}. Therefore, the strong size dependence of phonon lifetimes, along with reports of extended phonon coherence in layered perovskites, gives a promising outlook for long-lived and tunable coherent control of the quantum dynamics in perovskite nanocrystals.

\section{Methods}
{\bf Samples.}    The studied CsPbI$_3$ nanocrystals embedded in fluorophosphate Ba(PO$_3$)$_2$-AlF$_3$ glass were synthesized by rapid cooling of a glass melt enriched with the components needed for the perovskite crystallization. Details of the method are given in Refs.~\cite{Kolobkova2021,kirstein2023-SML}. The data on NC sizes evaluated from scanning transmission electron microscopy can be found in the supporting information section S5 in Ref.~\cite{Nestoklon2023}. Samples of fluorophosphate (FP) glass with the composition 35P$_2$O$_5$--35BaO--5AlF$_3$--10Ga$_2$O$_3$--10PbF$_2$--5Cs$_2$O (mol. \%) doped with BaI$_2$ were synthesized using the melt-quench technique. The glass synthesis was performed in a closed glassy carbon crucible at the temperature of $T=1050^\circ$C. Technological code of the studied sample is EK8. More information on its optical and spin-dependent properties can be found in Refs.~\onlinecite{Nestoklon2023} and~\onlinecite{Meliakov-2024NS}.  

The sample is placed in the variable temperature insert of a bath cryostat and kept in superfluid helium at the temperature of 2~K. The low energy exciton optical transitions in nanocrystals are located in the spectral range of 1.7--1.8~eV as follows from the photoluminescence spectrum, see Fig.~\ref{Fig1}(a). 

{\bf Photon echo.} Two-pulse photon echoes were measured using a transient four-wave mixing technique with heterodyne detection~\cite{Poltavtsev2016,Poltavtsev2020a,Poltavtsev2020,Poltavtsev2019}. We used a Ti:Sapphire laser as source of optical pulses with duration of $\sim100$~fs at the repetition rate of 75.75~MHz. All optical experiments were carried out at photon energies below 1.76~eV. The sample is excited resonantly by a sequence of two pulses separated in time by the delay $\tau_{12}$, giving rise to the formation of a four-wave mixing signal with electric field amplitude $\mathcal{E}_{\rm FWM}(t)$. The pulses are focused into a spot of about 200~$\mu$m diameter using a 0.5-m parabolic aluminum mirror. The incidence angles of the pulses are close to normal and equal to approximately $1/50$~rad and $2/50$~rad, corresponding to the in-plane wavevectors ${\bf k}_1$ and ${\bf k}_2$. An additional reference pulse is used for detection and its delay with respect to the first pulse is given by the time $\tau_{\rm ref}$. The delay times $\tau_{12}$ and $\tau_{\rm ref}$ are controlled using mechanical delay lines.  The four-wave mixing signal is measured in transmission geometry in the direction $\approx 3/50$~rad, which corresponds to the phase-matching direction $2{\bf k}_2 - {\bf k}_1$. 
Optical heterodyne detection is used to resolve the enhanced signal in time. By mixing with a relatively strong reference pulse and scanning $\tau_{\mathrm{ref}}$, the temporal profile of the photon echo pulse is measured.

Polarization sensitive measurements are performed by selecting proper polarization configurations (H-horizontal or V-vertical) for each of the beams using Glan-Thompson prism in conjunction with half-wave plates. In the detection path we also select the linear polarization by setting the polarization of the reference beam. We adopt a three-letter notation for the PE polarization configuration. The first two symbols correspond to the polarizations of the pulses exciting the sample, and the third symbol corresponds to the detection polarization. For example, in the HHH configuration, the polarizations of both excitation pulses and the detection are horizontal (H). In the HVH configuration, the second excitation pulse is vertically (V) polarized, while the first pulse and detection are horizontally polarized. HHH is accordingly termed co-linearly polarized ($\|$) and HVH cross-linearly polarized ($\times$) in Fig.~\ref{Fig:Osci}. We note that rotation of the sample does not lead to changes in the signal, i.e. the orientation of the sample is not important.

{\bf Raman spectroscopy. }
The Raman scattering signal was excited by a single-frequency Ti:Sapphire laser. The laser beam with the power of about 1~mW was focused on the sample to a spot with a diameter of about $300~\mu$m. The scattered light was analyzed by a Jobin-Yvon U1000 double monochromator with 1-meter focal length, allowing the high resolution of 0.2~cm$^{-1}$ (0.024 meV). The Raman signal was detected by a cooled GaAs photomultiplier and conventional photon-counting electronics. The Raman spectra were measured for co-polarized linear polarizations of excitation and detection.

\section{Acknowledgements}
The authors are thankful to I. A. Yugova and N. E. Kopteva for fruitful discussions. 
We acknowledge the financial support by the Deutsche Forschungsgemeinschaft: A.V.T., M.A.H, S.G., M.O.N and I.A.A (project AK40/13-1, no. 506623857) and D.R.Y. (project YA65/28-1, no. 527080192). The work of S.V.G. was supported by the NSF through DMR-2100248.
A.V.T, E.V.K. and M.S.K. acknowledge the Saint-Petersburg State University (Grant No. 125022803069-4).

\bibliography{perovskite_nano_crystals}

%% To extract SI, commment end document below, compile, extract SI by running 
%% $pdftk filename.pdf cat pS1-pS2 output filename_SI.pdf
%% then uncomment end document and run again to produce main file for submission.
%\end{document}

%=================================================
% Supplementary trick from https://tex.stackexchange.com/questions/168169/
\clearpage
\newpage
%\pagebreak
%\onecolumn
\onecolumngrid
\begin{center}
  \textsf{\textbf{\Large Supplementary Information:}}\\[0.2cm]
  \textbf{\large {\thetitle}}
  %{\textsf{\theauthors}\par}
  %{\textit{\theaffillist}\par}
\end{center}
\setcounter{equation}{0}
\setcounter{figure}{0}
\setcounter{table}{0}
\setcounter{page}{1}
\setcounter{section}{0}
\makeatletter
\renewcommand{\thepage}{S\arabic{page}}
\renewcommand{\theequation}{S\arabic{equation}}
\renewcommand{\thefigure}{S\arabic{figure}}
\renewcommand\theHfigure{Sfigure.\arabic{figure}}  % For hyperref not to confuse main figures and supplementary figures
\renewcommand{\thetable}{S\arabic{table}}
\renewcommand{\thesection}{S\arabic{section}}
\renewcommand{\thesubsection}{\thesection.\Alph{subsection}}
%\renewcommand{\bibnumfmt}[1]{[S#1]}
%\renewcommand{\citenumfont}[1]{S#1}
%\SectionNumbersOn
%\makeatletter
%\let\@startsection\acs@startsection
%\makeatother
%\renewcommand{\bibnumfmt}[1]{[S#1]}
%\renewcommand{\citenumfont}[1]{S#1}

%=================================================
%\refstepcounter{section} % To make hyperref use new internal numbers of eveerything

\newcommand{\tgo}{\gamma_0'}
\newcommand{\tge}{\gamma_1'}
\newcommand{\gph}{\gamma_{\rm ph}}

\section{Fine structure of exciton} \label{sec:si:FSS} % {SI:FS}

To take into account the fine structure of the exciton, we assume that the three-fold bright exciton state is split into mutually orthogonal linearly polarized components. For the photon echo calculations the following procedure is used: the dynamics is calculated in the eigenbasis of NC $\{\ket{0}$, $\ket{x'}$, $\ket{y'}$, $\ket{z'}\}$ then the equations for the density matrix components are rotated to the laboratory frame  $\{\ket{0}$, $\ket{x}$, $\ket{y}$, $\ket{z}\}$. The 1st pulse is assumed to be polarized along $z$ axis, the 2nd pulse at time $\tau$ is either polarized along $z$ or $x$ axis for HHH and HVH schemes respectively. The polarization along $z$ is detected at the photon echo time $2\tau$.
The orientation of a nanocrystal in the laboratory reference frame is described by three Euler angles $\alpha$, $\beta$, and $\gamma$.
The rotation matrix relating coordinates in $xyz$ and $x'y'z'$ bases is \cite{bookVarshalovich}: 
\newcommand{\eca}{c_{\alpha}}
\newcommand{\esa}{s_{\alpha}}
\newcommand{\ecb}{c_{\beta}}
\newcommand{\esb}{s_{\beta}}
\newcommand{\ecg}{c_{\gamma}}
\newcommand{\esg}{s_{\gamma}}
\begin{equation}\label{eq:rotmat}
  R(\alpha,\beta,\gamma) = \begin{pmatrix}
    \eca\ecb\ecg - \esa\esg & -\eca\ecb\esg-\esa\ecg & \eca\esb \\
    \esa\ecb\ecg + \eca\esg & -\esa\ecb\esg+\eca\ecg & \esa\esb \\
   -\esb\ecg                &  \esb\esg              & \ecb
  \end{pmatrix}\,,
\end{equation}
here and below in this section we use $s_{\xi} \equiv \sin\xi$, $c_{\xi} \equiv \cos\xi$ ($\xi=\alpha,\beta,\gamma$) to shorten the notation.

In the eigenbasis of NC, the dynamics is given by (we skip common factor $e^{-\gamma t}$):
\begin{align}\label{eq:3D_dyn_eigen}
   \rho_{0x'} &= e^{-t/T_2}e^{i\omega_1t}\rho_{0x'}^0\,,\;\;\;
 & \rho_{0y'} &= e^{-t/T_2}e^{i\omega_2t}\rho_{0y'}^0\,,\;\;\;
 & \rho_{0z'} &= e^{-t/T_2}e^{i\omega_3t}\rho_{0z'}^0\,;
\\
   \rho_{x'0} &= e^{-t/T_2}e^{-i\omega_1t}\rho_{x'0}^0\,,\;\;\;
 & \rho_{y'0} &= e^{-t/T_2}e^{-i\omega_2t}\rho_{y'0}^0\,,\;\;\;
 & \rho_{z'0} &= e^{-t/T_2}e^{-i\omega_3t}\rho_{z'0}^0\,.
\end{align}
Here $\hbar\omega_i$ ($i=1,2,3$) are the three exciton energies for excitons linerly polarized along $x'$, $y'$, $z'$ respectively. We assume that the diplole matrix element (and, consequently, radiative decay time) as well as the decoherence time $T_2$ is the same for all three exciton states.
After 1st pulse the components of density matrix are
\begin{equation}
  \rho_{0z}^{1+} = 1\,,\;\;\;
  \rho_{0y}^{1+} = \rho_{0x}^{1+} = 0\,.
\end{equation}
After 2nd pulse we have two options: for HHH scheme 
\begin{equation}
  \rho_{z0}^{2+} = \rho_{0z}^{2-}\,,\;\;\;\rho_{y0}^{2+} = \rho_{x0}^{2+} = 0\,.
\end{equation}
and for HVH scheme 
\begin{equation}
  \rho_{x0}^{2+} = \rho_{0x}^{2-}\,,\;\;\;\rho_{y0}^{2+} = \rho_{z0}^{2+} = 0\,.
\end{equation}
Rotating the dynamics equations \eqref{eq:3D_dyn_eigen} into laboratory coordinate frame using \eqref{eq:rotmat} gives rather lengthy equations for the amplitude of the signal: 
\begin{multline}
  A_{\|}(\alpha,\beta,\gamma) \propto \rho_{z0} 
 = \big[\eca^4\esb^4 + \esa^4\esb^4 + \ecb^4
\\
  + 2 \esa^2\eca^2\esb^4 \cos(\omega_1-\omega_2)\tau
  + 2 \eca^2\esb^2\ecb^2 \cos(\omega_1-\omega_3)\tau
  + 2 \esa^2\esb^2\ecb^2 \cos(\omega_2-\omega_3)\tau \big] e^{-2\tau/T_2}
\end{multline}
\begin{multline}
  A_{\times}(\alpha,\beta,\gamma) \propto \rho_{z0} 
  = \big[\eca^2\esb^2 (\eca\ecb\ecg-\esa\esg)^2 +
    \esa^2\esb^2 (\esa\ecb\ecg+\eca\esg)^2
  + \esb^2\ecb^2\ecg^2 
\\ 
  + 2 \esa\eca\esb^2 (\eca\ecb\ecg-\esa\esg)(\esa\ecb\ecg+\eca\esg) \cos(\omega_1-\omega_2)\tau
  - 2 \eca\esb^2\ecb\ecg(\eca\ecb\ecg-\esa\esg) \cos(\omega_1-\omega_3)\tau
\\
  - 2 \esa \esb^2\ecb\ecg (\esa\ecb\ecg+\esa\esg) \cos(\omega_2-\omega_3)\tau \big] e^{-2\tau/T_2}
\end{multline}
The result should be averaged over directions using:
\begin{equation}
  \left\langle A \right\rangle_{\rm ang} \equiv \frac1{8\pi^2} 
  \int_0^{2\pi} {\rm d} \alpha
  \int_0^{ \pi} \sin \beta {\rm d} \beta
  \int_0^{2\pi} {\rm d} \gamma \; A(\alpha,\beta,\gamma)
\end{equation}
Assuming the constant energies of three exciton components, but the arbitrary direction of the NCs, the amplitude of the echo is proporitional to 
\begin{align}
  \left\langle A_{\|} \right\rangle_{\rm ang} & \propto \left[ \frac65 + \frac4{15} \sum_{i>j} \cos(\omega_i-\omega_j)\tau \right] e^{-\frac{2\tau}{T_2}}\,,
\\
\left\langle A_{\times} \right\rangle_{\rm ang} & \propto \left[ \frac{2}{5} - \frac2{15} \sum_{i>j} \cos(\omega_i-\omega_j)\tau \right] e^{-\frac{2\tau}{T_2}}\,.
\end{align}
%Taking into account the decay of the levels which is assumed to be equal for all three exciton states, we end up with the final result
Now we also take into account the possible deviation of exciton energies. The absolute value of the exciton energy is not important for the phonon echo, only the splittings. E.g., the amplitude averaged over distribution of splittings between 1st and 2nd levels $\omega_{12} = \omega_1-\omega_2$ is given by
\begin{equation}
  \left\langle A \right\rangle_{\rm \omega_{12}} = \int W(\omega_{12}) A(\omega_{12},\omega_{13},\omega_{23}=\omega_{12}-\omega_{13}) d \omega_{12}
\end{equation}
Assuming the distribution function of the splittings to be Lorentzian, 
\begin{equation}
  W(\omega) = \frac{t_i^*}{\pi} \frac1{(\omega-\delta_i)^2 + \sqrt2/(t_i^*)^2}\,, 
\end{equation}
and that all three splittings are not correlated, the PE signal averaged over both the angles and the splittings distribution is given by 
% Use Gradstein Ryzhik 3.723 in page 424
\begin{align}
  \left\langle A_{\|} \right\rangle & \propto \left[ \frac{18}{15} + \frac4{15} \sum_{i=1,2,3} \cos(\delta_i \tau) e^{-\frac{2\tau}{t_i^*}} \right] e^{-\frac{2\tau}{T_2}}\,,
\\
\left\langle A_{\times} \right\rangle & \propto \left[ \frac{6}{15} - \frac2{15} \sum_{i=1,2,3} \cos(\delta_i \tau)  e^{-\frac{2\tau}{t_i^*}}  \right] e^{-\frac{2\tau}{T_2}}\,,
\end{align}
where the parameters $t_i^*$ work as an effective signal dephasing times and $\delta_i$ are the average splittings. An interesting result is that even though the PE signal is stable with respect to the level disorder, the disorder of its fine structure gives the contribution to the additional decay, not of the full signal, but of its oscillating components.
%\begin{equation}
%\mathrm{PE_{HHH}} = \left[ \frac{18}{15} + \sum_{i} \frac{4}{15} \cos\left( \frac{\delta_i\tau_{12}}{\hbar} \right) e^{-\frac{2\tau_{12}}{t_{i}^{*}}} \right] e^{-\frac{2\tau_{12}}{T_2}},
%\end{equation}
%\begin{equation}
%\mathrm{PE_{HVH}} = \left[ \frac{6}{15} - \sum_{i} \frac{2}{15} \cos\left( \frac{\delta_i\tau_{12}}{\hbar} \right) e^{-\frac{2\tau_{12}}{t_{i}^{*}}} \right] e^{-\frac{2\tau_{12}}{T_2}},
%\end{equation}

\section{Model of photon echo of exciton-polaron states.}\label{sec:PEcalcul}

In this section we outline the derivation of PE signal in the polaron model. We assume the exciton-polaron with relatively small Huang-Rhys factor which allows to perform calculations in four (six with account on light polarization) level scheme.

\subsection{Scheme of the levels and dynamics of the system}\label{sec:levels}

We consider the four-level system shown in Fig.~\ref{Fig:Scheme}b. %  Fig.~\ref{fig:fourlevel}.
There are: ground level $\ket{0}$, state with one phonon $\ket{0'}$, polaron state $\ket{X}$, and excited state of polaron $\ket{X'}$.
The energies of corresponding states are $0$, $\hbar\Omega$, $\omega_X$, $\omega_X+\hbar\Omega$.
In the calculations below we assume that spectral width of the laser pulse covers at least $\ket{X}$ and $\ket{X'}$ levels, this is in contrast to standard approximations for three-level atoms \cite{bookBerman}.
 
%%\begin{wrapfigure}[18]{r}{5.5cm}
%%\begin{tikzpicture}[>=Stealth]
%%  %\clip (-2.5,0) rectangle (2.5,6);
%%  %\draw[color=blue] (-2.5,0) rectangle (2.5,6);
%%  % Levels of exciton
%%  \draw[line width=1.5]  (-1.9,0.5) node[left]{$\ket{0}$}  -- (0.8,0.5);
%%  \draw[line width=1.5]  (-1.9,1.5) node[left]{$\ket{0'}$} -- (0.8,1.5);
%%  \draw[line width=1.5]  (-0.2,4.5) node[left]{$\ket{X}$} -- (2.5,4.5);
%%  \draw[line width=1.5]  (-0.2,5.5) node[left]{$\ket{X'}$}-- (2.5,5.5);
%%
%%  \draw[<-, >={Stealth},line width=1.0] (-1.0,1.5) -- ( 0.2,5.5) node [left,midway] {$\gamma_1$};
%%  \draw[<-, >={Stealth},line width=0.5] (-0.5,1.5) -- ( 0.4,4.5) node [left,midway]  {$\gamma'$};
%%  \draw[<-, >={Stealth},line width=1.0] (-0.5,0.5) -- ( 0.6,4.5) node [right,midway] {$\gamma_0$};
%%  \draw[<-, >={Stealth},line width=0.5] (-0.0,0.5) -- ( 1.6,5.5) node [right,midway] {$\gamma'$};
%%  %% phonon transitions
%%  \draw[<-, >={Stealth}, dashed] ( 0.5,0.5) -- ( 0.5,1.5) node [right,midway] {$\gamma_{ph}$};
%%  \draw[<-, >={Stealth}, dashed] ( 2.4,4.5) -- ( 2.4,5.5) node [left,midway] {$\gamma_{ph}$};
%%\end{tikzpicture}
%%\caption{Scheme of levels used in calculations.}\label{fig:fourlevel}
%%\end{wrapfigure}

Due to polaron formation, allowed transitions are not only $X \to 0$ (with probability $\gamma_0$) and $X'\to 0'$ (with probability $\gamma_1$), but also $X\to 0'$ and $X' \to 0$ (in this model, probability is the same $\gamma'$). In polaron model \cite{bookAPY}, the transition probability is proportional to product of dipole matrix element $|{\bf d}|^2$ and the overlap of oscillator functions shifted due to exciton formation which leads to extra factor 
\begin{subequations}\label{eq:gammaHR}
\begin{eqnarray}
  \gamma_0 & \propto & |{\bf d}|^2 e^{-2S_{HR}} \\
  \gamma'  & \propto & |{\bf d}|^2 S_{HR} e^{-2S_{HR}} \\
  \gamma_1 & \propto & |{\bf d}|^2 (1-S_{HR})^2 e^{-2S_{HR}}
\end{eqnarray}
\end{subequations}
where ${\bf d}$ is the dipole matrix elements of the optical transition, and we defined Huang-Rhys factor as $S_{HR}=\epsilon_{pol}/(\hbar\Omega)$ ($\epsilon_{pol}$ is the polaron formation energy).
Note that in this approximation polaron formation energy should be small which means that $\gamma' \ll \gamma_0, \gamma_1$.

In addition, we assume that the phonon may be emitted which leads to transitions $0'\to 0$ and $X' \to X$ with the same probability $\gamma_{ph}$.

To trace the polarization of the signal, we need to consider two polarization states of exciton, from now on  
the enumeration $0$ and $0'$ are ground state and state with phonon, $1(2)$ and $1'(2')$ are right(left) circularly polarized states of exciton(polaron) and first oscillator level of exciton(polaron), making the model effectively six-level. Still, we call the model four-level as here we neglect the splittings between levels (as the role of the fine structure is excluded, see Sec.~\ref{sec:si:FSS}) and we need it only to demonstrate that the phonon structure does not change the polarization structure of the signal. 
%Note that we take into account spin state of the exciton. 
Below, we extensively omit states $2$ and $2'$ when equations are same. 

The Lindblad equation, with all levels considered is:
\begin{equation}\label{eq:Lmain}
  \frac{\mathrm{d} \rho}{\mathrm{d} t} = -\frac{i}{\hbar} [H_0,\rho] + \mathcal{L}_d[\hat\rho] + \Gamma \odot \hat\rho\,.
\end{equation}
For convenience and compactness of the notation, with the use of $\odot$ for the Hadamard product (element-wise product or Schur product). 

The Hamiltonian is
\begin{equation}\label{eq:H0}
  H_0 = \begin{pmatrix} 
    0 & 0 & 0 & 0 & 0 & 0 \\
    0 & \hbar\Omega & 0 & 0 & 0 & 0 \\
    0 & 0 & \hbar\omega_X & 0 & 0 & 0\\
    0 & 0 & 0 & \hbar\omega_X & 0 & 0\\
    0 & 0 & 0 & 0 & \hbar\omega_X+\hbar\Omega & 0 \\
    0 & 0 & 0 & 0 & 0 & \hbar\omega_X+\hbar\Omega \\
  \end{pmatrix}
\end{equation}
and the ``inflow''diagonal part of superoperator is
\begin{equation}
  \mathcal{L}_d[\hat\rho] = \begin{pmatrix} 
    \gamma_{ph} \rho_{0'0'} + \gamma_0\rho_{X} + \gamma' \rho_{X'} & 0 & 0 & 0 & 0 & 0 \\
    0 & \gamma_1 \rho_{X'} + \gamma' \rho_{X} & 0 & 0 & 0 & 0 \\
    0 & 0 & \gamma_{ph}\rho_{1'1'} & 0 & 0 & 0 \\
    0 & 0 & 0 & \gamma_{ph}\rho_{2'2'} & 0 & 0 \\
    0 & 0 & 0 & 0& 0 & 0 \\
    0 & 0 & 0 & 0& 0 & 0
  \end{pmatrix}
\end{equation}
where for convenience we introduced $\rho_{X}=\rho_{11}+\rho_{22}$ and $\rho_{X'}=\rho_{1'1'}+\rho_{2'2'}$, and the matrix in the ``outflow'' part is
\begin{equation}
  \Gamma = - \frac12 \begin{pmatrix} 
    0    &  \gph      &      \tgo  &      \tgo &  \gph+\tge      &  \gph+\tge \\
    \gph & 2\gph      & \gph+\tgo  & \gph+\tgo & 2\gph+\tge      & 2\gph+\tge \\
    \tgo &  \gph+\tgo &     2\tgo &      2\tgo &  \gph+\tge+\tgo &  \gph+\tgo+\tge\\
    \tgo &  \gph+\tgo &     2\tgo &      2\tgo &  \gph+\tge+\tgo &  \gph+\tgo+\tge\\
    \gph+\tge &  2\gph+\tge & \gph+\tgo+\tge & \gph+\tgo+\tge &   2\gph+2\tge &  2\gph+2\tge\\
    \gph+\tge &  2\gph+\tge & \gph+\tgo+\tge & \gph+\tgo+\tge &   2\gph+2\tge &  2\gph+2\tge
  \end{pmatrix}
\end{equation}
where for convenience we defined 
\begin{equation}\label{eq:gammas_p}
  \gamma_0' = \gamma_0+\gamma'\,,\;\;\; \gamma_1' = \gamma_1+\gamma'\,.
\end{equation}

\subsection{Effect of the pulse on density matrix}\label{sec:DM_pulse}
Now we consider light-mediated transitions in the system shown in Fig.~\ref{Fig:Scheme}b. This is again generalization of the two-level system, check e.g. chapters 2, 6 and 10 of \cite{bookBerman}. 
The pulse is approximated by the plane wave with the smooth enveope ${\bf E}_0(z,t)$ and electric field given by
\begin{equation}\label{eq:Epulse}
  {\bf E}({\bf r},t) = {\bf E}_0 ({\bf k \cdot r}/k,t) e^{i({\bf k \cdot r}-\omega t)} + {\rm c.c.}
\end{equation}
Assuming the pulse duration is short compared with all coherence and relaxation times, the evolution of density matrix (DM) during the pulse may be found from the Schr\"odinger equation  
\begin{equation}\label{eq:rho_pulse}
  \dd{\rho}{t} = -\frac{\rm i}{\hbar} \left[H_0 + V , \rho \right]\,,
\end{equation}
where, in rotating wave approximation,
\begin{equation}\label{eq:V}
  V = \hbar \begin{pmatrix} 
    0 & 0 & \tilde{f}_+^* & \tilde{f}_-^* & s\cdot \tilde{f}_+^* & s \cdot \tilde{f}_+^* \\
    0 & 0 & s\cdot  \tilde{f}_+^* & s \cdot \tilde{f}_-^* & \tilde{f}f_+^* &  \tilde{f}_+^* \\
    \tilde{f}_+ & s\cdot \tilde{f}_+ & 0& 0& 0& 0\\
    \tilde{f}_- & s\cdot \tilde{f}_- & 0& 0& 0& 0\\
    s\cdot \tilde{f}_+ & \tilde{f}_+ & 0& 0& 0& 0\\
    s\cdot \tilde{f}_- & \tilde{f}_- & 0& 0& 0& 0\\
  \end{pmatrix}\,,
\end{equation}
\begin{equation}
  \tilde{f}_{\pm} = d E_0^{\pm} e^{-i\omega t} \equiv f_{\pm}e^{-i\omega t} \,,
\end{equation}
where $d$ is the dipole matrix element for the transitions $1,2 \to 0$, $s=\sqrt{S_{HR}}$, $E_0^{\pm}$ are circular components of the envelope \eqref{eq:Epulse}.

Solution of Eq.~\eqref{eq:rho_pulse} is surprisingly complex in general case, for two-level system it may be found in Ref.~\onlinecite{bookBerman}, but we are interested in four-level system and the case of small pulse area when the full solution is surplus. When the pulse area $\theta_{\pm}$ (equal to $f_{\pm}t_p$ for square pulse of duration $t_p$) is small compared with Rabi frequency, the first order correction to the density matrix may be obtained as 
\begin{equation}\label{eq:rho_pulse_1ord}
  \rho^+ = -\frac{i}{\hbar} \left[H_0 + V , \rho^- \right] t_p \equiv \hat{\mathcal{A}} \rho^-\,,
\end{equation}
where $\rho^-$ is the density matrix before pulse and $\rho^+$ is the density matrix after pulse.
Full solution of Eq.~\eqref{eq:rho_pulse_1ord} is too lengthy for our needs, below we will only give the results relevant for the calculations. 

For the PE we will need the action of the light pulse in the second order over pulse area. It may be shown that up to second order, it may be calculated as 
\begin{equation}\label{eq:rho_pulse_2ord}
  \rho^+ = \hat{\mathcal{A}} \hat{\mathcal{A}} \rho^-\,.
\end{equation}
Below we also give only part of this result relevant for photon echo calculations.

\subsection{Photon echo}
The signal in two circular polarizations is proportional \cite{bookBerman} to the corresponding components of polarization vector which is given by the following components of the density matrix: 
\begin{subequations}\label{eq:Ppm}
\begin{align}
  P_+ &\sim  \rho_{10 } + \rho_{1'0'} + s\rho_{1'0} + s \rho_{10'}\,,\\
  P_- &\sim  \rho_{20 } + \rho_{2'0'} + s\rho_{2'0} + s \rho_{20'}\,.
\end{align}
\end{subequations}
Note the extra factor $s$ in Eq.~\eqref{eq:Ppm} which originates from the ratio of diplole matrix elements for phonon-assisted transitions to the phonon-less transitions. Thus,
we are interested in the following components of the density matrix at times after second pulse: $\rho_{10}$, $\rho_{20}$, $\rho_{1'0}$, $\rho_{2'0}$, $\rho_{10'}$, $\rho_{20'}$, $\rho_{1'0'}$, $\rho_{2'0'}$.

\paragraph{Density matrix after 1st pulse}
The components of DM after 1st pulse are given by \eqref{eq:rho_pulse_1ord} which leads to:
\begin{equation}\label{eq:rho1_simpl}
  \rho_{01}^{1+} = i\theta_{1+}^*\,,\;\;\;  \rho_{01'}^{1+} = is\theta_{1+}^*\,, \;\;\;
  \rho_{02}^{1+} = i\theta_{1-}^*\,,\;\;\;  \rho_{02'}^{1+} = is\theta_{1-}^*\,. % ; \;\;\;
  %\rho_{j0}^{1+} = \left( \rho_{0j}^{1+} \right)^*\,. 
\end{equation}

\paragraph{Relaxation of DM components after 1st impulse}

One needs to trace the evolution of the following DM components: 
$\rho_{01}$, $\rho_{01'}$, $\rho_{02}$, $\rho_{02'}$. Below we give results only for right polarization, dynamics of other components is the same. 
Writing the components of the Lindblad equation \eqref{eq:Lmain}, we obtain the following dynamics of the DM components:
\begin{subequations}\label{eq:rho1p_fin}
\begin{align}
  \rho_{01 } \left( t\in (0,\tau_{12}) \right) &= \rho_{01}^{1+} e^{i\omega_Xt}e^{-\frac{\gamma_0'}2t}\,, \\
  \rho_{01'} \left( t\in (0,\tau_{12}) \right) &= \rho_{01'}^{1+} e^{i(\omega_X+\Omega)t}e^{-\frac{\gamma_1'+\gph}2t}\,, 
\end{align}
\end{subequations}
This leads to the following components of the DM at the time of 2nd pulse $t=\tau_{12}$
%\begin{subequations}
\begin{equation}\label{eq:PE:rho2b}
  \rho_{01 }^{2-} = \rho_{01 }^{1+} \varepsilon_X              \Gamma_0 \,,\;\;\;
  \rho_{01'}^{2-} = \rho_{01'}^{1+} \varepsilon_X \varepsilon  \Gamma_1\Gamma_p\,, 
%  \\
%  \rho_{02 }^{2-} &= \rho_{02 }^{1+} \varepsilon_X             \Gamma_0\,, & 
%  \rho_{02'}^{2-} &= \rho_{02'}^{1+} \varepsilon_X \varepsilon \Gamma_1\,, 
\end{equation}
%\end{subequations}
where for convenience we defined 
\begin{equation}
  \varepsilon_X = e^{i\omega_X \tau_{12}}\,,\;\; \varepsilon = e^{i\Omega \tau_{12}}\,,\;\;
  \Gamma_0 = e^{-\frac{\tgo}2 \tau_{12}}\,,\;\;\; 
  \Gamma_1 = e^{-\frac{\tge}2 \tau_{12}}\,,\;\;\; 
  \Gamma_p = e^{-\frac{\gph}2 \tau_{12}}\,.
\end{equation}
Note that $\Gamma_0$ and $\Gamma_1$ are almost equal and defined by the exciton coherence time (under our assumptions half of its lifetime). The difference between them is $\Gamma_1/\Gamma_0 =  e^{-\frac{\gamma_1-\gamma_0}2\tau_{12}} = e^{\frac{S_{HR}\gamma_0}2\tau_{12}}$ which is much smaller than all typical times in the system since $S_{HR}$ is small.

\paragraph{DM after 2nd pulse} From \eqref{eq:rho_pulse_2ord} we may compute the components relevant for the PE signal after the 2nd pulse:
\begin{subequations}\label{eq:rho2}
\begin{align}
  \rho_{10}^{2+} =  & \left[ \theta_{2+}^2(\rho_{01}^{2-}+s\rho_{01'}^{2-})
  +\theta_{2+}\theta_{2-}(\rho_{02}^{2-}+s\rho_{02'}^{2-}) \right]
\\
  \rho_{10'}^{2+} =   & \left[\theta_{2+}^2(s\rho_{01}^{2-}+ \rho_{01'}^{2-})
  +\theta_{2+}\theta_{2-}(s\rho_{02}^{2-}+\rho_{02'}^{2-}) \right]
\\
  \rho_{1'0}^{2+} = s & \left[ \theta_{2+}^2(\rho_{01}^{2-}+s\rho_{01'}^{2-})
  +\theta_{2+}\theta_{2-}(\rho_{02}^{2-}+s\rho_{02'}^{2-}) \right]
\\
  \rho_{1'0'}^{2+} = s & \left[ \theta_{2+}^2(\rho_{01'}^{2-}+s\rho_{01}^{2-})
  +\theta_{2+}\theta_{2-}(\rho_{02'}^{2-}+s\rho_{02}^{2-}) \right]
\end{align}
\end{subequations}
and
\begin{subequations}
\begin{align}
  \rho_{20}^{2+} = & \left[ \theta_{2+}\theta_{2-}(\rho_{01}^{2-}+s\rho_{01'}^{2-})
  +\theta_{2-}^2(\rho_{02}^{2-}+s\rho_{02'}^{2-}) \right] 
\\
  \rho_{2'0}^{2+} =   & \left[ \theta_{2+}\theta_{2-}(s\rho_{01}^{2-}+\rho_{01'}^{2-})
  +\theta_{2-}^2(s\rho_{02}^{2-}+\rho_{02'}^{2-}) \right] 
\\
  \rho_{20'}^{2+} = s & \left[ \theta_{2+}\theta_{2-}(\rho_{01}^{2-}+s\rho_{01'}^{2-})
  +\theta_{2-}^2(\rho_{02}^{2-}+s\rho_{02'}^{2-}) \right] 
\\
  \rho_{2'0'}^{2+} = s & \left[ \theta_{2+}\theta_{2-}(\rho_{01'}^{2-}+s\rho_{01}^{2-})
  +\theta_{2-}^2(\rho_{02'}^{2-}+s\rho_{02}^{2-}) \right] 
\end{align}
\end{subequations}

\paragraph{Relaxation of DM components after 2nd impulse} We we need to know the evolution of the DM components which enter \eqref{eq:Ppm}.
Below we give only components entering $P_+$. Dynamics of components entering $P_-$ is the same.

Writing the components of the Lindblad equation \eqref{eq:Lmain}, we obtain the following dynamics of the DM components:
\begin{subequations}\label{eq:rho2p_fin}
\begin{align}
  \rho_{10 } \left( t > t_2  \right) & = \rho_{10 }^{2+} e^{-i\omega_X(t-t_2)}e^{-\frac{\gamma_0'}2(t-t_2)}\,, \\
  \rho_{1'0} \left( t > t_2  \right) & = \rho_{1'0}^{2+} e^{-i(\omega_X+\Omega)(t-t_2)}e^{-\frac{\gamma_1'+\gamma_{ph}}2(t-t_2)}\,, \\
  \rho_{10'} \left( t > t_2  \right) & = \rho_{10'}^{2+} e^{-i(\omega_X-\Omega)(t-t_2)}e^{-\frac{\gamma_0'+\gamma_{ph}}2(t-t_2)}\,, \\
  \rho_{1'0'}\left( t > t_2  \right) & = \rho_{1'0'}^{2+}e^{-i\omega_X(t-t_2)}e^{-\frac{\gamma_1'+2\gamma_{ph}}2(t-t_2)}\,,
 \end{align}
\end{subequations}
Which gives the DM components at the time of echo signal $t-t_2 = \tau_{12}$:
\begin{subequations}\label{eq:PE:rho3b}
\begin{align}
  \rho_{10 }^{PE} &= \rho_{10 }^{2+} \varepsilon_X^*             \Gamma_0 \,, & 
  \rho_{1'0}^{PE} &= \rho_{1'0}^{2+} \varepsilon_X^*\varepsilon^*\Gamma_1\Gamma_p\,, 
  \\
  \rho_{1 0'}^{PE} &= \rho_{10'}^{2+}  \varepsilon_X^*\varepsilon \Gamma_0\Gamma_{p} \,, & 
  \rho_{1'0'}^{PE} &= \rho_{1'0'}^{2+} \varepsilon_X^*\Gamma_1\Gamma_{p}^2\,. 
  %\rho_{01 }^{2-} &= \rho_{01 }^{1+} \varepsilon_X              \Gamma_0 \,, & 
  %\rho_{01'}^{2-} &= \rho_{01'}^{1+} \varepsilon_X \varepsilon  \Gamma_1\,, 
  %\\
  %\rho_{02 }^{2-} &= \rho_{02 }^{1+} \varepsilon_X             \Gamma_0\,, & 
  %\rho_{02'}^{2-} &= \rho_{02'}^{1+} \varepsilon_X \varepsilon \Gamma_1\,, 
\end{align}
\end{subequations}
The evolution of components corresponding to second circular polarization is the same.

\paragraph{Contribution of different paths to PE}
To get the final result, we substitude \eqref{eq:rho1_simpl} into \eqref{eq:PE:rho2b} and then the result into \eqref{eq:rho2} and then from \eqref{eq:PE:rho3b} we get the DM at the time of PE. The amplitude of the signal in two polarizations is then obtained from Eqs.~\eqref{eq:Ppm}:
\begin{subequations}\label{eq:PE:Pp_from_rho3}
\begin{align}
  P_+ &\sim \rho_{10 }^{\rm PE} + \rho_{1'0'}^{\rm PE} + s\rho_{1'0}^{\rm PE} + s \rho_{10'}^{\rm PE}\,, \\
  P_- &\sim \rho_{20 }^{\rm PE} + \rho_{2'0'}^{\rm PE} + s\rho_{2'0}^{\rm PE} + s \rho_{20'}^{\rm PE}\,. 
\end{align}
\end{subequations}

Collecting all terms in the result we have for contributions to ${P}_{+}$ from $\rho_{10}$, $\rho_{1'0'}$, $ \rho_{1 0'}$, $\rho_{1'0 }$ originating from the DM components after 1st pulse given below in the beginning of each line: 
\begin{subequations}\label{eq:PE:Pp_half_res}
  \begin{align}
  \rho_{0 1 ,0 2 }\to \rho_{1 0 }^{\rm PE} & & & \Gamma_0^2 {A}_+\,,\\
  \rho_{0 1',0 2'}\to \rho_{1 0 }^{\rm PE} & & & s^2 \varepsilon \Gamma_0  \Gamma_1 \Gamma_p A_+ \,,\\
  \rho_{0 1 ,0 2 }\to \rho_{1 0'}^{\rm PE} & & & s^2 \varepsilon\Gamma_0^2 \Gamma_p A_+ \,,\\
  \rho_{0 1',0 2'}\to \rho_{1 0'}^{\rm PE} & & & s^2 \varepsilon^2 \Gamma_0\Gamma_1\Gamma_p^2 A_+ \,,\\
  \rho_{0 1 ,0 2 }\to \rho_{1'0 }^{\rm PE} & & & s^2 \varepsilon^* \Gamma_0\Gamma_1\Gamma_p A_+ \,,\\
  \rho_{0 1',0 2'}\to \rho_{1'0 }^{\rm PE} & & & s^4 \Gamma_1^2\Gamma_p^2 A_+ \,,  \label{eq:PE:Pp_half_res_to_drop}\\
  \rho_{0 1 ,0 2 }\to \rho_{1'0'}^{\rm PE} & & & s^2 \Gamma_0\Gamma_1\Gamma_p^2 A_+ \,,\\
  \rho_{0 1',0 2'}\to \rho_{1'0'}^{\rm PE} & & & s^2 \varepsilon \Gamma_1^2\Gamma_p^3 A_+ \,,
\end{align}\end{subequations}
where
\begin{subequations}
\begin{equation}
  A_+ = \theta_{2+} \left( \theta_{2+} \theta_{1+}^* +  \theta_{2-} \theta_{1-}^* \right)\,.
\end{equation}
Note that the term \eqref{eq:PE:Pp_half_res_to_drop} may be dropped as it exceeds precision of equations.

The contribution to ${P}_{-}$ from $\rho_{20}$, $\rho_{2'0'}$, $ \rho_{2 0'}$, $\rho_{2'0 }$ is the same as \eqref{eq:PE:Pp_half_res} but instead of $A_+$ the polarization dependence is given by 
\begin{equation}
  A_- = \theta_{2-} \left( \theta_{2-} \theta_{1-}^* +  \theta_{2+}\theta_{1+}^* \right)\,.
\end{equation}
\end{subequations}

It is convenient to write both as 
\begin{equation}
  A_{\pm} = \theta_{2\pm} A_0\,,\;\;\; A_0 = \theta_{2-} \theta_{1-}^* +  \theta_{2+}\theta_{1+}^*\,,
\end{equation}
to see that the polarization dependence of the PE signal coincides with second impulse polarization while its amplitude depends on respective polarization of first and second pulses.

As a result, the amplitude of the non-oscillating signal is 
\begin{equation}\label{eq:PE:Pfin0}
  \mathcal{P}_{\pm}^{PE,0} \sim \left[ \Gamma_0^2 + S_{HR} \Gamma_0\Gamma_1\Gamma_p^2 \right] A_{\pm} \,.
\end{equation}
The amplitude of the signal oscillating at $\Omega$ is 
\begin{equation}\label{eq:PE:Pfin1}
  \mathcal{P}_{\pm}^{PE,\Omega} \sim S_{HR} \Gamma_p \left[\Gamma_0\Gamma_1\left( \varepsilon + \varepsilon^* \right) 
  + \varepsilon \left( \Gamma_0^2 + \Gamma_1^2\Gamma_p^2 \right) \right] A_{\pm} \,.
\end{equation}
and of the signal oscillating at $2\Omega$ is 
\begin{equation}\label{eq:PE:Pfin2}
  \mathcal{P}_{\pm}^{PE,2\Omega} \sim S_{HR} \Gamma_0 \Gamma_1 \Gamma_p^2 \varepsilon^2 A_{\pm}  \,.
\end{equation}

\paragraph{Result}

Let us first consider the general scheme of the polarization dependence of PE signal. Following~\cite{Poltavtsev2019}, we consider linearly polarized excitation pulses with second pulse rotated to angle $\phi$ and linearly polarized detection with polarized rotated to $\phi_d$. Then 
\begin{equation}
  \theta_{1+} = \theta_{1-} = 1/\sqrt2 \,,\,\,\,
  \theta_{2+} = e^{i\phi}/\sqrt2\,,\,\,\, \theta_{2-} = e^{-i\phi}/\sqrt2\,.
\end{equation}
Which gives $A_0 = \cos{\phi}$ and 
\begin{equation}
  A_{\pm} = \frac{e^{\pm i \phi}}{\sqrt2} \cos{\phi}\,.
\end{equation}
Amplitude of linearly-polarized heterodyne detection of main PE signal is proportional to
\begin{equation}\label{eq:PE:Pd0}
  P_d^{0} = 
  \left\vert \frac{e^{i(\phi-\phi_d)} + e^{-i(\phi_d-\phi)} }2 \cos{\phi} \right\vert
  = \left\vert \cos{(\phi-\phi_d)} \cos{\phi} \right\vert
\end{equation}
We have two configurations: HHH and HVH. First is $\phi=\phi_d=0$ and second is $\phi_d=0$, $\phi=\pi/2$. This gives 
\begin{equation}
  P_d^{0,HHH} = 1\,,\;\;\; P_d^{0,HVH} = 0.
\end{equation}

As long as polarization dependence on the excitation/detection angles \eqref{eq:PE:Pd0} is real (which is the case for HHH and HVH configurations), the absolute value of product is the product of absolute values and signal is zero for HVH scheme and for HHH scheme (we omit index HHH below) is:
\begin{equation}\label{eq:PE:Pd_prel}
  \mathcal{P}_{d}^{PE} =\Gamma_0^2  \left\vert 
  \left[ 1 + S_{HR} \widetilde{\Gamma}\Gamma_p^2 \right] 
  + S_{HR} \Gamma_p \left[\widetilde{\Gamma}\left( \varepsilon + \varepsilon^* \right)  + \varepsilon (1+\widetilde{\Gamma}^2\Gamma_p^2) \right]
  +  S_{HR} \widetilde{\Gamma}\Gamma_p^2 \varepsilon^2 \right\vert 
\end{equation}
where we introduced 
\begin{equation}
  \widetilde{\Gamma}  = \Gamma_1/\Gamma_0= e^{S_{HR} \gamma_0\tau_{12}}\,.
\end{equation}
Note that $\widetilde{\Gamma}$ is $~1$ as compared with both, $\Gamma_0$ and $\Gamma_p$.

To write explicitly the result for the PE amplitude, we use the fact that for $t \ll 1$ and arbitrary complex $\epsilon$:
\begin{equation}\label{eq:complexabs}
  \left\vert 1 + t \epsilon \right\vert = 1 + t \mathrm{Re} (\epsilon)
  + \frac{t^2}{2} \left(  \mathrm{Im} (\epsilon) \right)^2 + O \left( t^3 \right)\,.
\end{equation}
We keep second order terms to highlight the fact that even for single frequency of the complex amplitude, the absolute value in \eqref{eq:PE:Pd_prel} will result in second order in $S_{HR}$ signal at double frequency. 

Using \eqref{eq:complexabs} we may write the final result (keeping only terms to the first order in $S_{HR}$)
\begin{equation}\label{eq:PE:Pd}
  \mathcal{P}_{d}^{PE} =\Gamma_0^2 
  \left\{ 1 + S_{HR}\Gamma_p  \left[
  \widetilde{\Gamma}\Gamma_p
  + \left(1 + 2\widetilde{\Gamma} + \widetilde{\Gamma}^2\Gamma_p^2 \right)\cos(\Omega \tau_{12}) 
  + \widetilde{\Gamma}\Gamma_p  \cos(2 \Omega \tau_{12})
  \right] \right\}
\end{equation}

Or, explicitly, 
\begin{multline}\label{eq:PE:Pdfin}
  \mathcal{P}_{d}^{PE} = e^{-\gamma_0'\tau_{12}}
  \Bigg[ 1 + S_{HR} e^{-\frac{\gph}2\tau_{12}} \Bigg(   
  e^{-\frac{\gph-2S_{HR}\gamma_0}2\tau_{12}}
  \\
+  \left(1 + 2e^{S_{HR} \gamma_0\tau_{12}} + e^{-(\gph-2S_{HR} \gamma_0)\tau_{12}} \right) \cos(\Omega \tau_{12}) 
  % \\
+ e^{-\frac{\gph-2S_{HR}\gamma_0}2\tau_{12}} \cos(2 \Omega \tau_{12})
\Bigg) \Bigg]\,,
\end{multline}
and, finally, with the help of Eqs.(\ref{eq:gammaHR},\ref{eq:gammas_p}): 
\begin{multline}\label{eq:PE:Pdfin2}
  \mathcal{P}_{d}^{PE} = e^{-(1+S_{HR})\gamma_0\tau_{12}}
  \Bigg[ 1 + S_{HR} e^{-({\gph}-S_{HR}\gamma_0)\tau_{12}} \\ 
  + S_{HR} \cos(\Omega \tau_{12})e^{-\frac{\gph-2S_{HR} \gamma_0}2\tau_{12}}   
 \left(e^{-S_{HR} \gamma_0\tau_{12}} + 2 + e^{-\gph\tau_{12}} \right)  
  \\
 + S_{HR}\cos(2 \Omega \tau_{12})e^{-({\gph}-S_{HR}\gamma_0)\tau_{12}} 
 \Bigg]\equiv \Psi_0(\tau_{12};\Omega)\,.
\end{multline}

\end{document}